\documentclass[showpacs, pra,twocolumn,preprintnumbers ,amsmath, amssymb, superscriptaddress, aps]{revtex4-2}
\usepackage{color}
\usepackage{amsmath,amssymb}
\usepackage{pifont}% http://ctan.org/pkg/pifont
\usepackage{amssymb}  % More symbols
\usepackage{bbold}
\usepackage{float}
\usepackage{subfloat}
\usepackage{adjustbox}

\usepackage[caption=false]{subfig}
\usepackage{tikz}
\usepackage{makecell}
\usepackage{subfig}
\usepackage{pifont}   % Ding symbols
\usepackage{graphicx} % Include figure files
\graphicspath{{Figuress/}}
\usepackage{dcolumn}  % Align table columns on decimal point
\usepackage{bm}       % bold math
\usepackage{multirow} % Table functions
\usepackage{placeins}% For making floats not move around everywhere
\usepackage[colorlinks]{hyperref}
\usepackage{mathtools}
\usepackage{appendix}

\newcommand{\bb} {\color{blue}}

\def \be{\begin{align}}
	\def \ee{\end{align}}
\def \bea{\begin{eqnarray}}
	\def \eea{\end{eqnarray}}

\begin{document}	
	\title{Tunneling effect of fermions in silicene through  {potential barrier}}
	\date{\today}
	
	\author{Sanae Zriouel}
	%\email{sanae.zriouel@usms.ma}
	\affiliation{Faculty of Sciences and Technology, CADI AYYAD University, Marrakech. Morocco}
	
	\author{Ahmed Jellal}
	\email{a.jellal@ucd.ac.ma}
	\affiliation{Laboratory of Theoretical Physics, Faculty of Sciences, Choua\"ib Doukkali University, PO Box 20, 24000 El Jadida, Morocco}
	\affiliation{Canadian Quantum  Research Center,
		204-3002 32 Ave Vernon, Vernon BC V1T 2L7,  Canada}

	\pacs{ 73.22.Pr, 72.80.Vp, 73.63.-b\\
		{\sc Keywords}: Silicene, rectangular barrier, Dirac equation,
		transmission, conductance, Klein effect, Fabry-P\'{e}rot interference.
	}

\begin{abstract}

The influence of  a rectangular potential barrier on  the quantum transport of fermions in silicene is explored. 
%We explore how a rectangular {potential barrier} influences the quantum transport of fermions in silicene. 
Specifically, analytical solutions are presented to derive transmission and reflection probabilities together with conductance. It is shown that the transmission is highly sensitive to both the barrier height and incident energy. As a result,  the occurrence of Klein and resonant tunnelings is observed, with a significant dependence on the barrier width. Notably, it is found that perfect transmission extends beyond normal incidence, occurring at various oblique angles. Moreover, the transmission pattern exhibits a more fragmented structure with increasing barrier width, reminiscent of Fabry-P\'{e}rot resonances. In contrast, the conductance displays a non-monotonic dependence on incident energy and features rapid oscillations with a rising barrier height. However, at a constant barrier height, there is a minimal disparity among conductance profiles for high incident energy values. When incident energy equals the barrier height, the conductance experiences a local minimum. For a thin barrier,  a substantial reduction in conductance is observed, unlike the oscillatory behavior seen with a thicker barrier. These findings underscore the progress in silicene research and offer a fresh perspective on the relativistic applications of tunneling in this material.

\end{abstract}

\maketitle

%%%%%%%%%%%%%%%%%%%%%%%%%%%%%%%%%%%%%%%%%%%

\section{Introduction}

%%%%%%%%%%%%%%%%%%%%%%%%%%%%%%%%%%%%%%%%%

Since the discovery of graphene \cite{1} and the incredible progress made in this field of study, significant efforts have been made to theoretically and empirically explore comparable 2D materials. This emerging field of study, which encompasses applications such as transistors \cite{Schwierz,Das} and topological field-effect transistors {(FETs)} \cite{Liu,Qian}, as well as fundamental physics \cite{Castro} and chemistry \cite{Yuan,Kalantar}, could have a significant impact on science and technology. Due to its unusual features and technological compatibility, silicene, a counterpart of silicon for graphene, has recently gained intense interest among 2D materials \cite{Kara,Xu}. It is worth noting that this last was reported for the first time by Takeda and Shiraishi in $1994$ \cite{Takeda} and then, Guzman-Verri {\it et al.} \cite{Guzman} re-examined it and gave it the name silicene in $2007$. However, neither silicene nor a solid phase of silicon that resembles graphite appear to occur in nature. The exfoliation techniques used to make graphene cannot be used to make pure  silicene layers.

%%%%%%%%%%%%%%%%%%%%%%%%%%%%%%%%%%%%%%%%%
Silicene differs significantly from graphene despite sharing many of its
properties, including the honeycomb lattice structure and relativistic Dirac dispersion. First of all, the 2D lattice of silicene is buckled due to the enormous atomic radius of silicon. To put it another way, in contrast to graphene, which has a genuine single atomic layer structure, silicene has two atomic layers that are spaced apart perpendicularly by a distance of $d\approx 0.46$~\AA  \ \cite{6,7} as illustrated in Fig. \ref{fig1}a. This buckled configuration leads the silicon atoms to bond in a $sp^{2}$-$sp^{3}$ mixed hybridization rather than a pure $sp^{2}$ hybridization \cite{Pulci} {as for graphene}. Furthermore, pristine silicene has a tunable bandgap of 1.55 meV \cite{Feng} and a rather significant spin-orbit coupling gap of 3.9 meV \cite{Jiang}. These properties can be changed by adjusting the
on-site potential  using an external electric field \cite{6,7}. 
When {this field} is increased, the silicene exhibits a topological phase transition from a quantum spin Hall state to the band insulator. Topologically non-trivial electronic states bring a rich physics into play and may even enable the preservation of quantum information in devices made of these materials \cite{Tahir,Kim}. From an experimental point of view, it is essential to succeed in the synthesis of practical silicene. %On some substrates, 
Recent experimental studies have grown certain silicene phases epitaxially \cite{Aufray,Lalmi,Dybala}. 
It has also been possible to create silicene heterostructures using hexagonal boron nitride \cite{Wiggers} and silicene multilayers intercalated with graphite surfaces \cite{Kupchak}. Because of the intense interaction with metallic substrates, the electronic band structure of silicene undergoes unfortunate and noteworthy changes \cite{Cheng,Cahangirov}. 
A significant solution is also needed for the limited lifetime of silicene, which is caused by both its structural instability and
the possibility of oxidation from exposure to the environment \cite%
{Solonenko}. These challenges prevent there from being a widely used method for synthesising silicene. However, some intriguing approaches have been made, such as the passivation \cite{Ritter} or encapsulation \cite{Tao} processes, which benefit electronic applications, like the fabrication of  {FETs}. 
Silicene-based FETs working at room temperature have just recently been created, which addresses the long-standing issues of oxidation and thermodynamic instability and represents a substantial step toward practical applications of silicene \cite{Akinwande}. It has already been proven that silicene FETs can display ambipolar behavior with mobility of carriers of  100 cm$^{2}$/Vs at room temperature \cite{Tao}. 
This has given silicene new possibilities for use in functional device applications.

We investigate the impact of a rectangular {potential barrier} on the quantum transport of fermions in silicene. Specifically, we present  analytical solutions for deriving transmission and reflection probabilities, as well as conductance. Our findings reveal that the transmission is remarkably sensitive to both the barrier parameters (height, width) and incident energy. This led to the observation of Klein and resonant tunnelings. Noteworthy is the discovery of the perfect transmission occurs not only at normal incidence but also at various oblique angles. Additionally, the transmission pattern demonstrates an increasingly fragmented structure with a growing barrier width, reminiscent of Fabry-P\'{e}rot resonances. In contrast, the conductance displays rapid oscillations with an increasing barrier height. It also exhibits a non-monotonic dependence on incident energy. However, a minimal difference is found in conductance profiles for high incident energies and a given barrier height. When incident energy equals the barrier height, the conductance experiences a local minimum. Compared to the oscillatory behavior seen with a thicker barrier, a notable reduction in conductance is noticed with a thin barrier.

%%%%%%%%%%%%%%%%%%%%%%%%%%%%%%%%%%%%%%%%%
The structure of this paper is as follows: In Sec. \ref{MTM}, we will delineate the technical aspects of our methodology for computing quantum transport {properties} using a Dirac-like Hamiltonian.  Specifically, we will address the incorporation of different components into the Hamiltonian system that describes particles scattered by a rectangular silicene barrier. Subsequently, we will determine the energy spectrum for each region, expressed in terms of various scattering parameters. In Sec. \ref{TPB}, we will solve the scattering problem by employing boundary conditions to explicitly derive transmission and reflection probabilities. We will present numerical results, discuss their implications, and compare them with other theoretical and experimental findings from the literature. In Sec. \ref{CON}, the associated conductance will be theoretically computed using the Landauer-B\"{u}ttiker formalism and numerically analyzed. Finally, we summarize and conclude our work.

%%%%%%%%%%%%%%%%%%%%%%%%%%%%%%%%%%

\section{Theoretical model}\label{MTM}

%%%%%%%%%%%%%%%%%%%%%%%%%%%%%%%%%%%%%

\begin{figure*}[tbph]
\begin{center}
\includegraphics[scale=0.25]{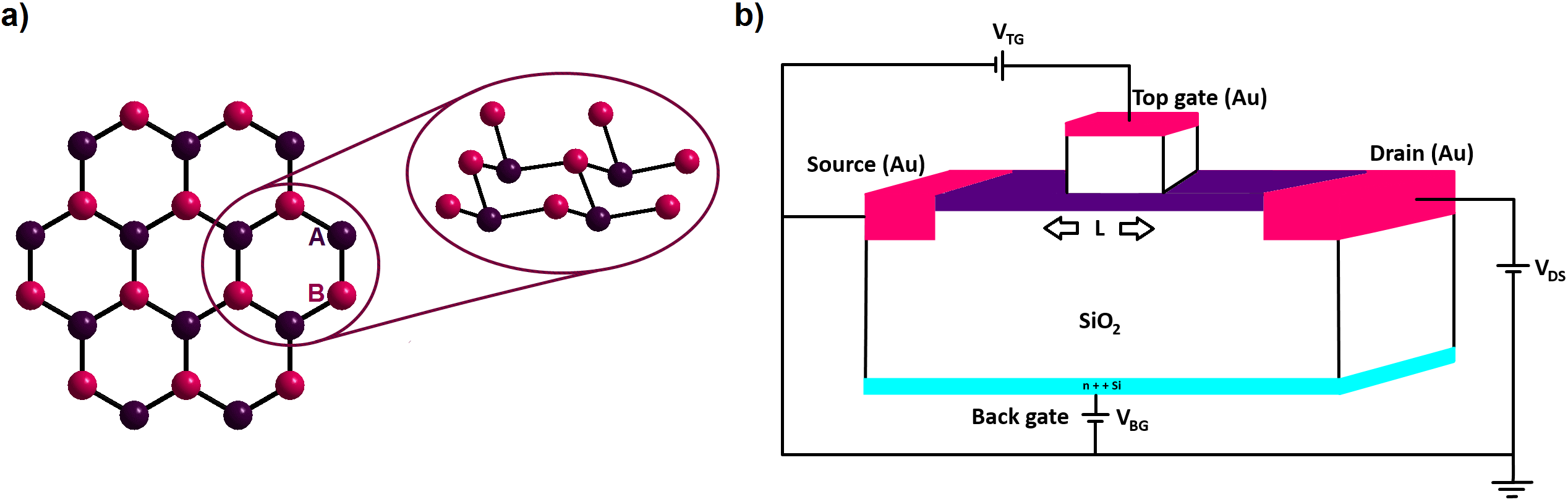}
\end{center}
\par
\vspace{-10pt}
\caption{(color online) a): The crystal structure of silicene is a 2D honeycomb lattice where the ionic size of the silicon atoms causes the A and B sublatices to sit in separate parallel planes. b): Design of a device illustrating our system.  }
\label{fig1}
\end{figure*}

\begin{figure*}[tbph]
	\begin{center}
		\includegraphics[scale=0.7]{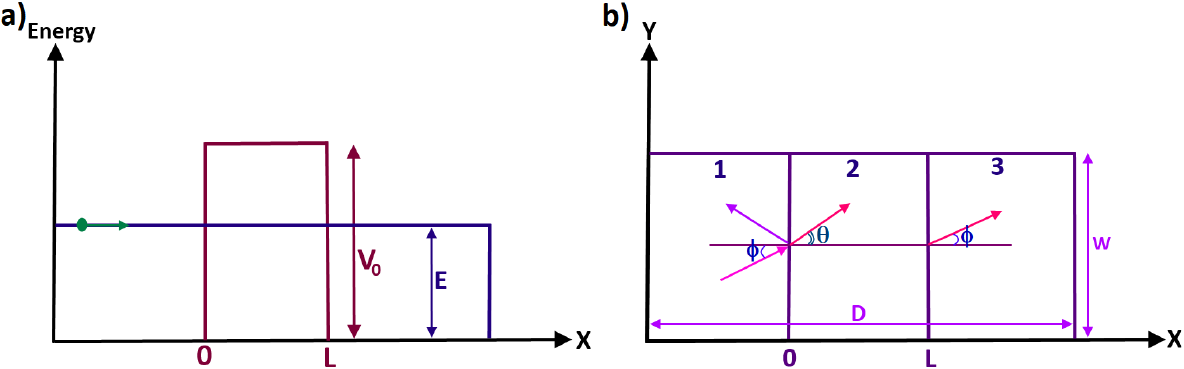}
	\end{center}
	\par
	\vspace{-10pt}
	\caption{(color online) a): Visualization of the  {potential barrier \(V(x)\) of height $V_0$ and width $L$} responsible for scattering Dirac fermions in silicene. b): A depiction of the angles \(\phi\) and \(\theta\) associated with the scattering formalism within the three regions, denoted as 1, 2, and 3. The distances \(D\) and \(W\) represent the dimensions of the silicene sheet.}
	\label{fig2}
\end{figure*}

As depicted in Fig. \ref{fig1}a, silicene is a 2D honeycomb lattice and has a crystal structure. Because of the ionic size of silicon atoms, the A and B sublattices take positions in distinct parallel planes. In Fig. \ref{fig1}b, we consider a device design contrasted with a rectangular single layer of silicene that is deposited on the dielectric layer SiO$ 2$ and has two gold contacts for the source and drain, a back gate, and a top gate \cite{10}. The top gate modifies the current that flows through the device from source to drain, passing over a  {potential barrier}, while the rear gate regulates the charge carrier density of the sample. Since the sample is large enough to not be considered a strip, we assume that there are no edge effects. As seen in Fig. \ref{fig2}a along the $x$-direction, we take into account a silicene sheet that is subject to a square  {potential barrier}. According to Fig. \ref{fig2}b, particles strike the barrier from one side at an angle $\phi$ before forming an angle $\theta$ following transmission. As a result, the present system is composed of three regions labeled by $j=1, 2, 3$, 
which can be described by a  Hamiltonian at low-energy excitations around Dirac points $\eta =\pm$, at the hexagonal first Brillouin zone corners $K$ and $K'$. {This is} 
\begin{equation}
H= v_{F}\left( \sigma _{x}p_{x}-\eta \sigma _{y}p_{y}\right) -\eta \tau
_{z}\triangle _{so}\sigma _{z}+V\left( x\right) \mathbb{I}_{2}  \label{1}
\end{equation}%
where $v_F$ is the Fermi velocity, $\sigma_i\ (i=x, y, z) $ are the Pauli matrices, $\vec p =(p_x, p_y) $ is the momentum. 
 The second term in \eqref{1} represents the Kane-Mele term for the intrinsic spin-orbit coupling \cite{5}, with $\Delta _{so}=3.9$ meV {that is  a distinctive feature of silicene},
 and the index $\tau _{z}$ refers to the two spin degrees of freedom, up $\left( \tau _{z}=+\right) $ and down $\left(\tau _{z}=-\right) $. The third term is associated with the rectangular potential barrier
  of width $L$ and height $V_{0}$ (see Fig. \ref{fig2}a)
\begin{equation}
V\left( x\right) =V_{j}=\left\{ 
\begin{array}{ccc}
0,\text{ \ } & x<0\qquad  & \text{(region }1\text{)} \\ 
V_{0}, & 0<x<L & \text{(region }2\text{)} \\ 
0,\text{ \ } & x>L\qquad  & \text{(region }3\text{)}%
\end{array}%
\right.   \label{2}
\end{equation}
and $\mathbb{I}_2$ is the $2\times 2$ unit matrix. Since the commutation relation $[H, p_y] =0$ holds, we can separate variables and write the eigenspinors as plane waves in the $y-$direction, i.e., $\psi_{j}(x,y)=e^{ik_{y}y}(\phi _{j}^{+}(x),\phi _{j}^{-}(x))^{\bb T}$, with ${\bb T}$ denoting the transpose.

At this point, we move forward by addressing the eigenvalue equation $H\psi _{j}= E\psi _{j}$ for the valley $\eta $ and spin $\tau _{z}$ within each region. Specifically, in region $1$, the eigenspinors can be expressed in relation to the incident and reflected waves as 
\begin{widetext}
\begin{equation}
\psi _{1}(x,y)=\frac{e^{ik_{y}y}}{\sqrt{DW\left( 1+\alpha ^{2}\right) }}%
\left[  
\begin{pmatrix}
1 \\ 
s_{1}\alpha e^{-i\eta \phi }%
\end{pmatrix} e^{ik_{x}x}+r
\begin{pmatrix}
1 \\ 
-s_{1}\alpha e^{i\eta \phi }%
\end{pmatrix} e^{-ik_{x}x}\right]  \label{4}
\end{equation}
\end{widetext} 
where {$D$ and $W$ denote the dimensions of  the silicene sheet along the $x$- and $y$-directions, respectively. Here,} $r$ is the reflection coefficient, 
$
	\alpha =\sqrt{\frac{E_{1}+\eta \tau _{z}\triangle _{so}}{E_{1}-\eta \tau
			_{z}\triangle _{so}}} $
			% \label{5}
%\end{equation}
and $\phi =\arctan (k_{y}/k_{x})$ is the angle that incident particles make with $x$-direction, see Fig. \ref{fig2}b.
The sign function $s_{1}=\text{sgn}(E_{1}-\eta \tau
_{z}\triangle _{so})$ refers to the conduction and valence bands.
The $x$- and $y$-components of the particle wave vector can be easily obtained
\begin{equation}
k_{x}=k
%\frac{\sqrt{E_{1}^{2}-\Delta _{so}^{2}}}{\hbar v_{F}}
\cos \phi
,\quad k_{y}= k
%\frac{\sqrt{E_{1}^{2}-\Delta _{so}^{2}}}{\hbar v _{F}}%
\sin \phi  \label{6}
\end{equation}%
showing the dispersion relation 
\begin{equation}
	E_1= s_1\hbar v_F \sqrt{k^2 + \Delta _{so}^{2}}
%k=\frac{\sqrt{E_{1}^{2}-\Delta _{so}^{2}}}{\hbar \upsilon _{F}}  \label{7}
\end{equation}%
with {$k=|\vec k|$ and  $\vec k=(k_x,k_y)$ is the wave vector}.
In region $2$, the solution gives the eigenspinors 
\begin{widetext}
\begin{equation}
\psi _{2}(x,y)=\frac{e^{ik_{y}y}}{\sqrt{DW\left( 1+\beta ^{2}\right) }}\left[
c
\begin{pmatrix}
1 \\ 
s_{2}\beta e^{-i\eta \theta }%
\end{pmatrix} e^{iq_{x}x}+d
\begin{pmatrix}
1 \\ 
-s_{2}\beta e^{i\eta \theta }%
\end{pmatrix} e^{-iq_{x}x}\right]  \label{8}
\end{equation}%
\end{widetext}
where $(c,d)$ are two constants, 
$	\beta =\sqrt{\frac{E_{2}-V_{0}+\eta \tau _{z}\triangle _{so}}{%
			E_{2}-V_{0}-\eta \tau _{z}\triangle _{so}}}$
and $\theta =\arctan (k_{y}/q_{x})$ such as {the two components}
\begin{align}
q_{x}=q
%\frac{\sqrt{(E_{2}-V_{0})^{2}-\Delta _{so}^{2}}}{\hbar v _{F}}%
	\cos \theta, \quad
	& k_{y}=q
	%\frac{\sqrt{(E_{2}-V_{0})^{2}-\Delta _{so}^{2}}}{%
%		\hbar v _{F}}
		\sin \theta  \label{9}
\end{align}
{are} associated to the wave vector $\vec q=(q_x, k_y)$ with  {$q=|\vec q|$}
\begin{align}
q=\frac{\sqrt{(E_{2}-V_{0})^{2}-\Delta _{so}^{2}}}{\hbar v _{F}}
\end{align}
leading to the energy $E_2$.
{Here,} we have defined $s_{2}=\text{sgn}(E_{2}-\eta \tau
_{z}\triangle _{so}-V_{0})$. 
Finally in region $3$, the transmitted wave is 
\begin{equation}
\psi _{3}(x,y)=\frac{te^{ik_{y}y}}{\sqrt{DW\left( 1+\alpha ^{2}\right) }}% 
\begin{pmatrix}
1   \\ 
s_{1}\alpha e^{-i\eta \phi}
\end{pmatrix} e^{ik_{x}x}  \label{10}
\end{equation}%
{where} $t$ is the transmission coefficient that will be explicitly determined in the next. 

Conversely, given the nature of the system, energy conservation must be satisfied, ensuring \(E = E_{1} = E_{2} = E_{3}\). Additionally, based on the {transverse momentum conservation $k_y$ in \eqref{6} and \eqref{9}, we establish a relation between the incident angle $\phi$ and the angle $\theta$ obtained in region 2, which is} 
\begin{equation}
\sin \theta =\sqrt{\frac{E^{2}-\Delta _{so}^{2}}{(E-V_{0})^{2}-\Delta
_{so}^{2}}}\sin \phi .  \label{111}
\end{equation}%
We note that there exists a maximum angle of incidence where electrons are  transmitted at {$\frac{\pi}{2}$} for a given energy. Beyond this angle, the \(x\)-component of the wave vector {($k_x$)} becomes imaginary. The incident plane waves transform into evanescent waves as the electrons traverse the barrier, and real exponents take place. 
The condition {(realized by injecting $\theta=\frac{\pi}{2}$ into \eqref{111})} establishes the limit angle
\begin{equation}
\phi _{l}=\arcsin \sqrt{1+\frac{V_{0}^{2}-2V_{0}E}{E^{2}-\Delta _{so}^{2}}}.
\label{12}
\end{equation}%
%%%%%%%%%%%%%%%%%%%%%%%%%%%%%%%%%%%%%%%%%%%%%%%%%%%%%%%%%
\begin{figure*}[tbph]
	\begin{center}
		\includegraphics[scale=0.65]{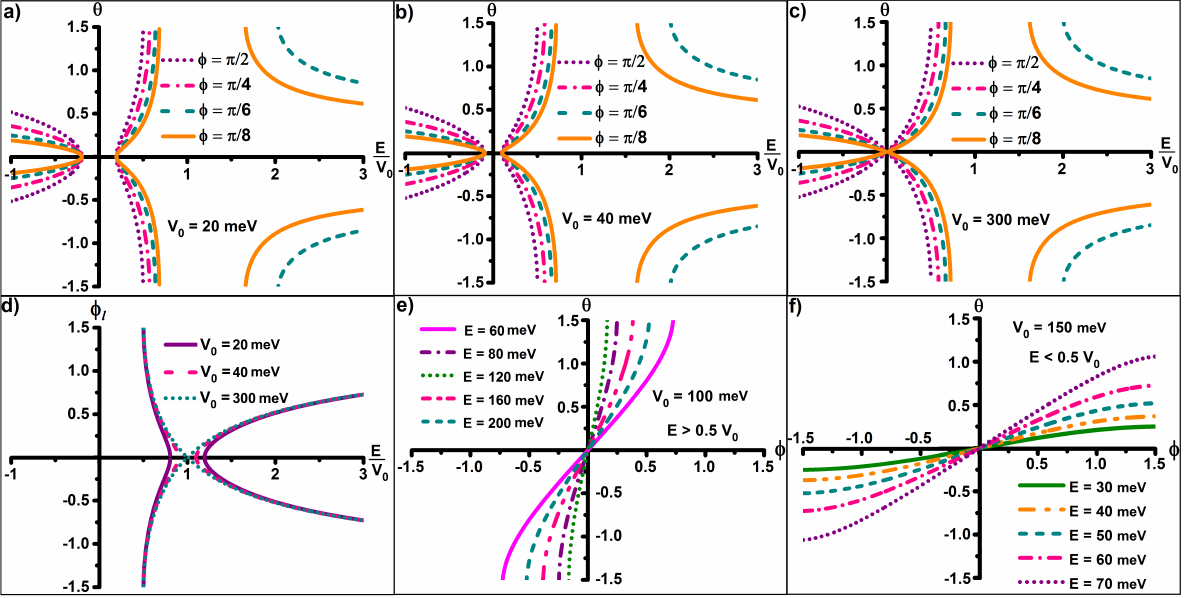}
	\end{center}
	\par
	\vspace{-10pt}
	\caption{(color online) a,b,c): The angle $\theta$ versus the energy $E$
		for various values of the incident angle $\phi $ and barrier height $V_{0}$.
		d): Limited  incident angle $\phi_l$ versus the energy for different {values of} $V_0$. e,f): The  angle $\theta$
		versus  the  incident angle  $\phi $ for different {values of} $E$.}
	\label{fig3}
\end{figure*}
Complete reflection at the barrier is not feasible for electrons with energies $E\in ]-\infty ,-\Delta _{SO}[\cup]V_{0}+\Delta _{SO},+\infty \lbrack $. Consequently, these electrons undergo transmission with an angle smaller than the incident one, approaching the normal direction. Electrons with higher energy  can be transmitted using a plane wave only if the incident angle falls within the range \(-\phi_{l}\) to \(\phi_{l}\), corresponding to their energy value. Electrons with an energy equal to \(0.5V_0\) are transmitted with an angle equal to the incident one, achieved by deviating from their usual direction. The limiting angle is zero for the energy \(E=V_{0}\pm \Delta_{SO}\). Specifically, only plane waves with normal incidence can successfully pass through the barrier.
In a strict sense, Fig. \ref{fig3} illustrates the behavior of emerged angles $\theta$ and $\phi$, {as described by \eqref{111}}. Specifically, Figs. \ref{fig3}(a, b, c) depict the  angle $\theta$ as a function of the energy $E$ for barrier height: \(V_0 = 20\) meV, 40 meV, 300 meV, and incident angle: \(\phi = \frac{\pi}{6}\), \(\frac{\pi}{4}\), \(\frac{\pi}{2}\). Figs. \ref{fig3}(e, f) illustrate the  angle $\theta$ versus the incident angle $\phi$ for fixed energy values in two cases: \(V_{0} = 100\) meV (\(E > 0.5V_{0}\)) and \(V_{0} = 150\) meV (\(E < 0.5V_{0}\)). Fig. \ref{fig3}(d) displays the limited incident angle \(\phi_{l}\) as a function of the energy $E$ for \(V_{0} = 20\) meV, 40 meV, 300 meV. It is noteworthy that for large barrier heights, our results exhibit similarities to those obtained for monolayer graphene \cite{8}.

%\vspace{2mm}
\section{Transmission }\label{TPB}

%%%%%%%%%%%%%%%%%%%%%%%%%%%%%%%%%%%%%%%%%%%%%%%%%%%%%%%%%%%
We will conduct a comprehensive analysis of the transmission probability for Dirac fermions in silicene when scattered by a {potential barrier}  based on the results obtained previously. Our focus will be on exploring the impact of structural disorder arising from the width and height of the barrier on transmission. Additionally, we will delve into various strategies for manipulating transport properties. To start, let us use the continuity of eigenspinors \eqref{4}, \eqref{8}, and \eqref{10} at the two interfaces $x=0$ and $x=L$.
After a lengthy algebra, we can solve the obtained set of equations to end up with the transmission $t$ and reflection $r$ coefficients established as
\begin{widetext}
\begin{eqnarray}
t &=&\frac{\left( 1+e^{2\eta i\theta }\right) \left( 1+e^{2\eta i\phi
}\right) e^{-iL\left( k_{x}-q_{x}\right) }}{1+e^{2\eta i\left( \theta +\phi
\right) }+e^{2i\left( \eta \theta +q_{x}L\right) }+e^{2i\left( \eta \phi
+q_{x}L\right) }+\frac{s_{1}s_{2}\left( \alpha ^{2}+\beta ^{2}\right) \left[
e^{i\eta \left( \theta +\phi \right) }-e^{i\left( \eta \left( \theta +\phi
\right) +2q_{x}L\right) }\right] }{\alpha \beta }}  \label{14} \\
r &=&\frac{e^{-i\eta \phi }\left( -1+e^{2iq_{x}L}\right) \left[ -\alpha
^{2}e^{i\eta \theta }+s_{1}s_{2}\alpha \beta \left( e^{i\eta \phi }-e^{i\eta
\left( 2\theta +\phi \right) }\right) +\beta ^{2}e^{i\eta \left( \theta
+2\phi \right) }\right] }{s_{1}s_{2}\alpha \beta \left[ 1+e^{i\eta 2\left(
\theta +\phi \right) }+e^{i2\left( \eta \theta +q_{x}L\right) }+e^{i2\left(
q_{x}L+\eta \phi \right) }\right] +\left( \alpha ^{2}+\beta ^{2}\right) %
\left[ e^{i\eta \left( \theta +\phi \right) }-e^{i\left( \eta \left( \theta
+\phi \right) +2q_{x}L\right) }\right] }  \label{15}.
\end{eqnarray}%
\end{widetext}
Subsequently, by using the relations  {\(T = tt^{\ast}=|t|^2\) and \(R = rr^{\ast}=|r|^2\)}, we derive the transmission and reflection probabilities for electrons traversing the {potential barrier}  in silicene. {They are given by}
\begin{widetext}
\begin{eqnarray}
T &=&\frac{\cos ^{2}\theta \cos ^{2}\phi }{\left[ \cos \theta \cos \phi \cos
\left( q_{x}L\right) \right] ^{2}+\sin^{2} \left( q_{x}L\right) \left[ \frac{%
\alpha ^{2}+\beta ^{2}}{2\alpha \beta }-s_{1}s_{2}\sin \theta \sin \phi %
\right] ^{2}}  \label{16} \\
R &=&\frac{\frac{\alpha ^{2}+\beta ^{2}}{\alpha \beta }\left[ \frac{\alpha
^{2}+\beta ^{2}}{4\alpha \beta }-s_{1}s_{2}\sin \theta \sin \phi \right]
+\sin^{2} \theta -\cos^{2} \phi }{\left[ \frac{\cos \theta \cos \phi }{\tan
\left( q_{x}L\right) }\right] ^{2}+\left[ \frac{\alpha ^{2}+\beta ^{2}}{%
2\alpha \beta }-s_{1}s_{2}\sin \theta \sin \phi \right] ^{2}}.  \label{17}
\end{eqnarray}
\end{widetext}
At this level, we have some comments in order. The barrier is transparent to all the electrons impinging perpendicularly on it, regardless of their energy \cite{11}. This phenomenon, known as the Klein paradox \cite{12}, is absent in non-relativistic electrons, where the transmission coefficient at normal incidence is consistently less than one. The barrier is also transparent for electrons with $q_{x}L=n\pi$ {as can be seen in \eqref{16}, with $n\in \mathbb{Z}$}, resulting in a relationship between energy and incident angle 
\begin{equation}
\left( \frac{\pi \hbar \upsilon _{F}}{L}\right) ^{2}n^{2}=V_{0}\left(
V_{0}-2E\right) +\left( E^{2}-\Delta _{so}^{2}\right) \cos ^{2}\phi 
\label{18}
\end{equation}
{which can be established  by using \eqref{9} together with \eqref{111}.}
We made the assumption that the angle at which electrons approach the barrier is smaller than the limit angle $\phi_l$ \eqref{12}. If they approach at a larger angle, the analysis remains valid, and the only modification needed is to replace \(q_x\) with \(i q_x\), transforming plane waves \(e^{\pm i q_{x}x}\) into evanescent waves \(e^{\mp i q_{x}x}\). In this scenario, the electron wavefunction comprises exponential terms outside the barrier without any oscillatory components.
Regarding the coefficients, we have to replace $\cos\left(q_{x}L\right) $ by $\cosh \left( q_{x}L\right) $ and $\sin \left( q_{x}L\right) $ by $i\sinh\left( q_{x}L\right) $. In particular, in order to analyze the transport properties, it is interesting to express the transmission coefficient as a function of {the energy $E$, barrier parameters $(V_0,L)$, and  incident angle $\phi$}. By requiring the condition on angles $\phi \leq \phi _{l}$, $\phi \geq \phi _{l}$ and $\phi=0$, we can distinguish three different cases. These give the following transmissions in order:
\begin{widetext}
\begin{align}
&T(E,V_{0},L,\phi )=\frac{1}{1+V_{0}^{2}\frac{\left( \left( E^{2}-\Delta
_{so}^{2}\right) \tan ^{2}\phi +\frac{\Delta _{so}^{2}}{\cos ^{2}\phi }%
\right) \sin ^{2}\left( \frac{L}{\hbar \upsilon _{F}}\sqrt{V_{0}\left(
V_{0}-2E\right) +\left( E^{2}-\Delta _{so}^{2}\right) \cos ^{2}\phi }\right) 
}{\left( E^{2}-\Delta _{so}^{2}\right) \left( V_{0}\left( V_{0}-2E\right)
+\left( E^{2}-\Delta _{so}^{2}\right) \cos ^{2}\phi \right) }} \label{19}\\
&T(E,V_{0},L,\phi )=\frac{1}{1+V_{0}^{2}\frac{\left( \left( E^{2}-\Delta
_{so}^{2}\right) \tan ^{2}\phi +\frac{\Delta _{so}^{2}}{\cos ^{2}\phi }%
\right) \sinh ^{2}\left( \frac{L}{\hbar \upsilon _{F}}\sqrt{V_{0}\left(
2E-V_{0}\right) +\left( \Delta _{so}^{2}-E^{2}\right) \cos ^{2}\phi }\right) 
}{\left( E^{2}-\Delta _{so}^{2}\right) \left( V_{0}\left( 2E-V_{0}\right)
+\left( \Delta _{so}^{2}-E^{2}\right) \cos ^{2}\phi \right) }} \label{20}\\
&T(E,V_{0},L,0)=\left( 1+V_{0}^{2}\frac{\Delta _{so}^{2}\sin ^{2}\left( \frac{%
L}{\hbar \upsilon _{F}}\sqrt{\left( E-V_{0}\right) ^{2}-\Delta _{so}^{2}}%
\right) }{\left( E^{2}-\Delta _{so}^{2}\right) \left( \left( E-V_{0}\right)
^{2}-\Delta _{so}^{2}\right) }\right) ^{-1}.  \label{21}
\end{align}
\end{widetext}
As we can see, the energy $E$,  barrier parameters ($V_0, L$), and incident angle $\phi $ all affect the transmission, but not silicene sheet dimensions ($D, W$). Additionally, with regard to the incident angle $\phi$, the transmission  is symmetric. It is important to note that  \eqref{21} anticipates complete transmission under conditions of normal incidence \((\phi \rightarrow 0, \theta \rightarrow 0)\). The Klein paradox for Dirac particles falls under that category \cite{11,Katsnelson,Beenakker}. The electron then immediately penetrates the barrier without reflecting. This phenomenon is known as a transmission resonance, and it is unaffected by the magnitude of $\phi$. Moreover, we note that \(T(0)\approx 1\) is satisfied for any of \(q_x L\) values under normal incidence, indicating complete transparency of the barrier.

\begin{figure*}[tbph]
\begin{center}
\includegraphics[width=17.cm,height=11cm]{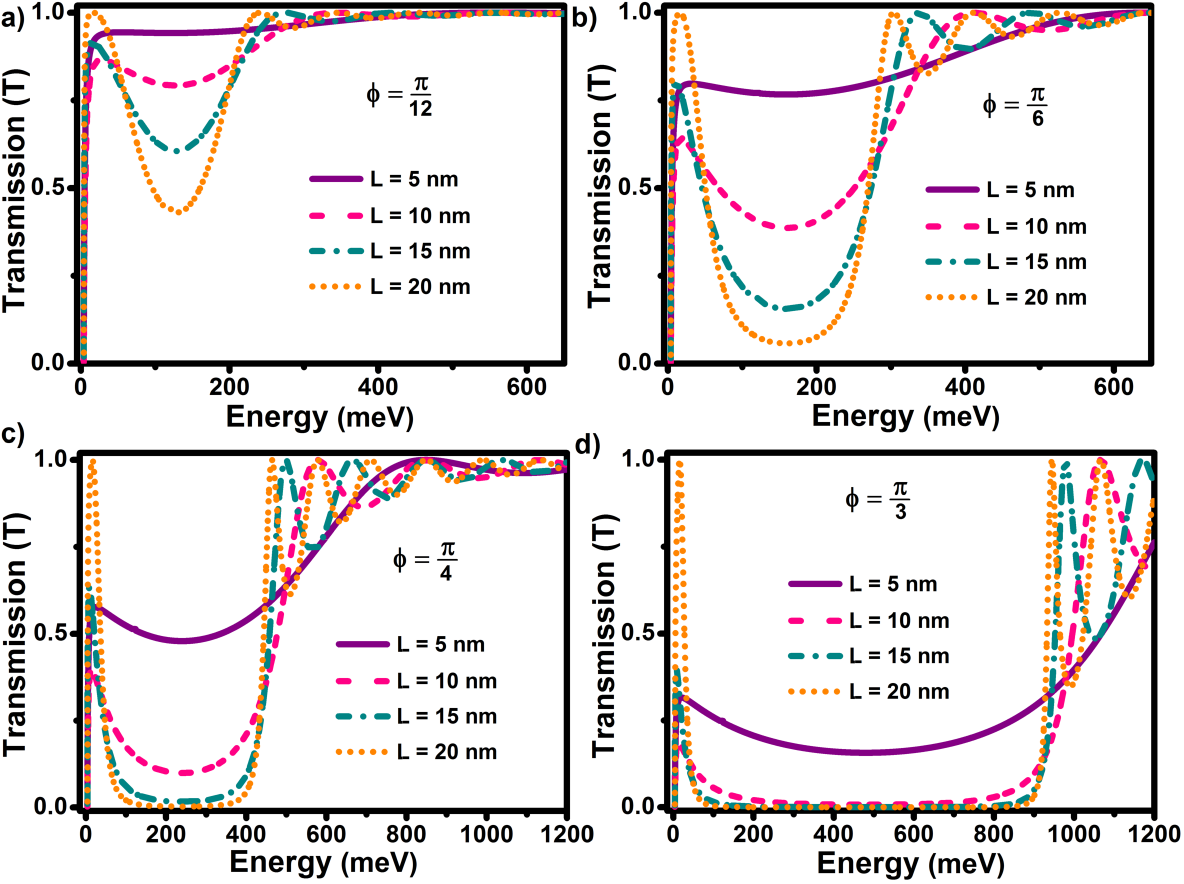}
\end{center}
\par
\vspace{-10pt}
\caption{(color online) Transmission probability $T$  versus the energy  $E$ for  barrier height $%
V_{0}=120$ meV and different values of incident angle $\phi$.
%: a) $\frac{\protect\pi }{12}$, b) 
%$\frac{\protect\pi }{6}$, c) $\frac{\protect\pi 
%}{4}$, and d) $\frac{\protect\pi }{3}$.
}
\label{fig4}
\end{figure*}

In what follows, we provide numerical results for various transport properties and tunneling effect of Dirac fermions in silicene subjected to a rectangular {potential barrier}.  
Fig. \ref{fig4} depicts the transmission probability $T$ as a function of energy $E$ for a fixed incident angle and barrier width $L$ ranging from 5 nm to 20 nm. Overall, the transmission is oscillating with a minimum for electrons of energies close to the barrier height $V_ 0 $.  
However, there is an exception resulting from $L=5$ nm for $\phi =\frac{\pi }{12}$ that exhibits an exponential rise with the increase of   energy.
The number of resonant tunneling peaks increases as the barrier width $L$ increases for a fixed $\phi $. Furthermore, the width of the resonant tunneling peaks observed in the transmission  decreases with increasing $L$, which means that the valence band quasi-bound state gets thinner with the increase in  $L$.  Moreover, the presence of resonant tunneling peaks in the transmission  suggests that the electrons can resonantly tunnel through the hole-like quasi-bound state lying within the valence band inside the barrier. The transmission  exhibits some backscattering at $\phi =\frac{\pi }{12}$ and $\frac{\pi }{6}$, which is more pronounced in the energy around $V_0$, where no Klein tunneling occurs.  
It is demonstrated that when $L\leq 15$ nm for $\phi =\frac{\pi }{4}$\ and $L=5$ nm for $\phi =\frac{\pi }{3}$, the transmission spectra are very similar to those of  $\phi =\frac{\pi }{12}$\ and $\frac{\pi }{6}$. In contrast, the transmission spectra for all other $L$ values show a transmission gap around $V_0$ that grows with $L$. In fact, the transmission between the energy levels,  $-\sqrt{(V_{0}\tan \phi)^{2}\cos ^{-2}\phi +\Delta _{so}^{2}}+V_{0}\cos ^{-2}\phi<E<
\sqrt{(V_{0}\tan \phi )^{2}\cos ^{-2}\phi+\Delta _{so}^{2}}+V_{0}\cos ^{-2}\phi $,
%and $-\sqrt{(V_{0}\tan \phi)^{2}\cos ^{-2}\phi +\Delta _{so}^{2}}+V_{0}\cos ^{-2}\phi $, 
is becoming closer to zero. In the case where the incident angle is greater than the limit angle, transmission takes place via evanescent waves. Furthermore, as the barrier width $L$ (or incident angle $\phi$) decreases (increases), the transmission gap around $V_0$ decreases (increases).

\begin{figure*}[tbph]
\begin{center}
\includegraphics[width=17.cm,height=11cm]{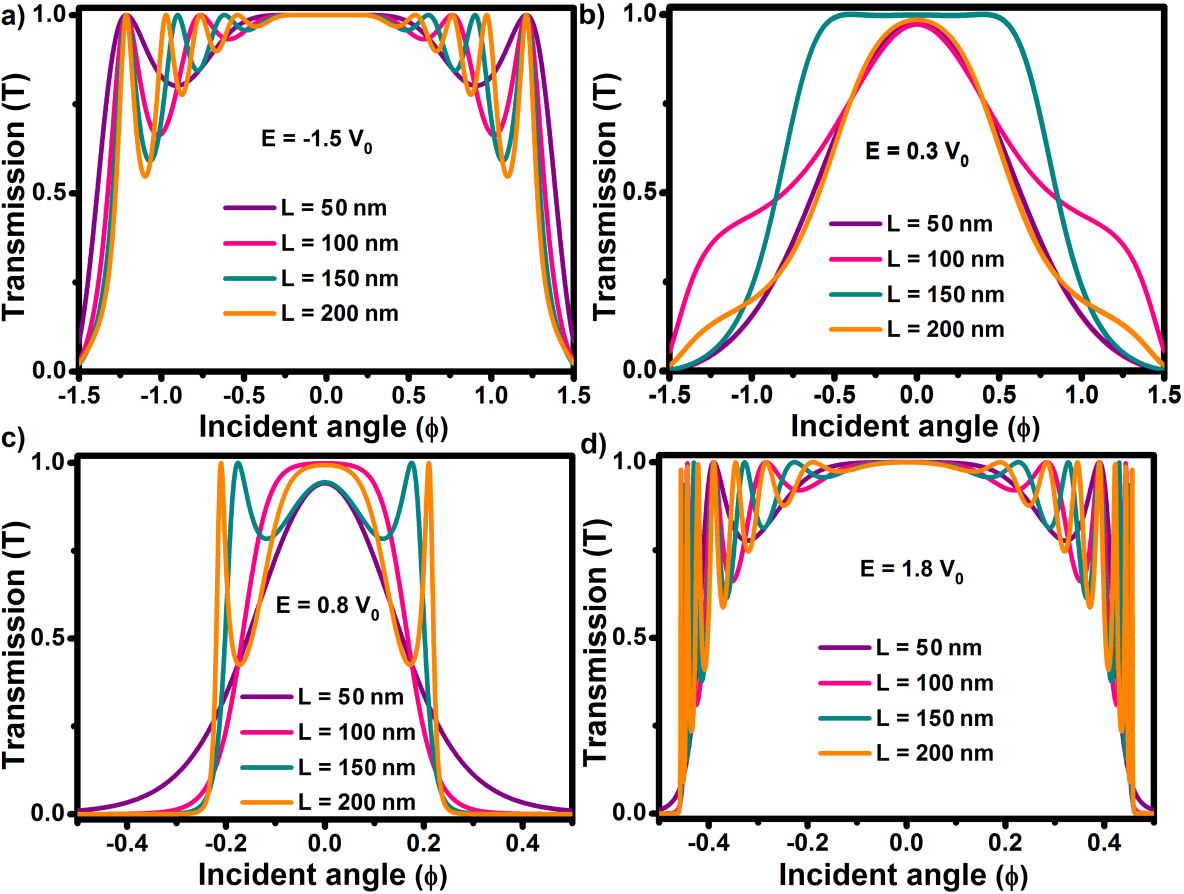}
\end{center}
\par
\vspace{-10pt}
\caption{(color online) Transmission probability $T$ versus the incident angle $\phi$
for barrier height $V_{0}=100$ meV, different values  of energy $E$ and barrier width $L$.}
%(a) $%
%E=-150$ meV, (b) $E=30$ meV, (c) $E=80$ meV, and (d) $E=180$ meV.}
\label{fig5}
\end{figure*}
{The angular dependence of transmittance $T$ is plotted in Fig. \ref{fig5}
 by varying the incident angle $\phi $ and fixing the energy $E$.}
For comparison, transmission probabilities for different barrier widths $L=50$ nm, $100$ nm, $150$ nm, and $200$ nm are depicted. The evolution of the transmission spectrum clearly distinguishes two transport modes based on the energy $ E $: one for Figs. \ref{fig5}(a,c) with $-\Delta _{so}<E<V_{0}+\Delta _{so}$ and the other for  Fig. \ref{fig5}d with $E>V_{0}+\Delta _{so}$. All the figures show that at the normal incidence $\phi =0$, the transmission for both transport schemes approaches unity. Then, the barrier is completely transparent, which is a manifestation of perfect tunneling when the incident wave vector is normal to the interface. Moreover, when the incident angle $\phi $ increases,  $T$ tends to decrease more or less steeply. Figs. \ref{fig5}(a,b) show that transmission is allowed over the entire range of incident angle $\left( -1.5\leqslant \phi \leqslant 1.5\right) $, whereas Fig. \ref{fig5}c shows that transmission is limited below the lower critical angle and above the upper critical angle, which depends on barrier width $L$. Beyond the critical angle, evanescent waves dominate, and transmission is essentially nonexistent. However, as shown in Fig. \ref{fig5}d, there are noticeable transmission resonances at certain incident angles where $T$ tends to unity. 
%We investigated how the barrier width $L$ affected the transmission and 
{Additionally, we discover} that as $L$ increases, $T$ shrinks symmetrically in the $\phi$-space. The resonances described in  \eqref{18} begin to appear when $ L $ is large enough.

\begin{figure*}[tbph]
\begin{center}
\includegraphics[width=17.cm,height=11cm]{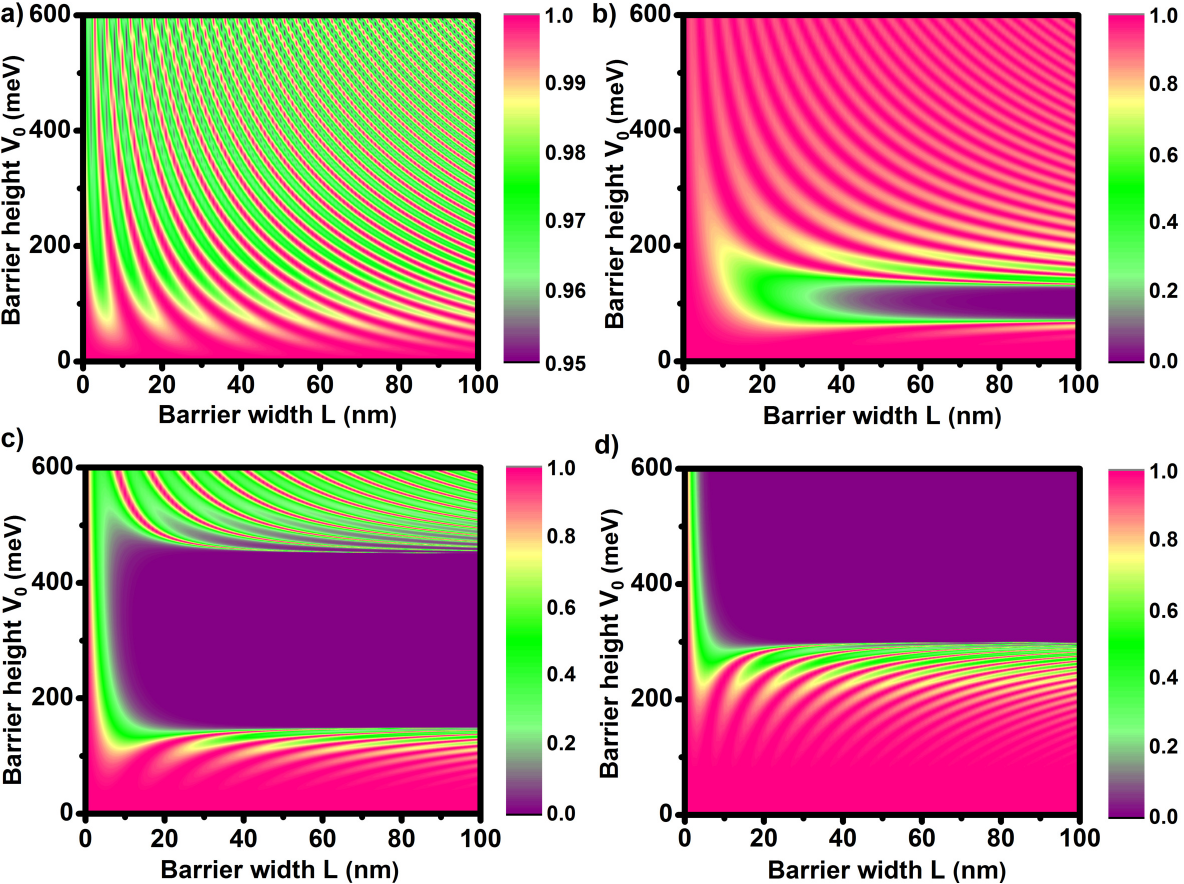}
\end{center}
\par
\vspace{-10pt}
\caption{(color online) Density plot of transmission probability $T$ as a function of the barrier height $V_{0}$ and width $L$ for different value of energy $E$ and incident angle $\phi$: 
%	. $L$(nm) are represented on the horizontal axis and $V_{0}$(eV) on the vertical one. 
	a) $E=-100$ meV and $\protect\phi =\frac{\pi}{12}$, b) $E=100$ meV and $\protect\phi =\frac{\pi}{12}$, c) $E=300$ meV and $\protect\phi =\frac{\pi}{6}$,  d) $E=600$ meV and $\protect\phi =\frac{\pi}{6}$.}
\label{fig6}
\end{figure*}

Fig. \ref{fig6} shows how the barrier height $V_0$ and width $L$ affect the transmission $T$, where each hue corresponds to a specific value of $T$ from bottom to top. In the presence of a {potential barrier such that}  $E<V_0$, $T$ approaches the unity regardless of width $L$, a characteristic of Klein tunneling. Furthermore, the dependence of $T$ on $V_ 0 $ and $L$ reveals that it oscillates between large and small values as a function of $L$. 
Due to quantum interference occurring within the barrier for \(E < V_0\), the transmission displays a series of Fabry-Pérot resonances. As the energy transitions from negative to positive, the tunneling behavior undergoes significant changes, leading to an intricate transmission distribution. First, we observe that for \(V_{0} < \frac{E}{2}\), the transmission is maximal and remains constant regardless of the barrier width $L$. In the case where \(\frac{E}{2} < V_{0} < \frac{3E}{2}\), the combined influence of both energy and barrier width on transmission becomes predominant. With an increase in barrier width $L$, the transmission gradually decreases until it reaches zero at a critical barrier width, a value dependent on either the incident angle or the energy. Consequently, a new low transmission gap emerges, indicating a nearly perfect reflection of incident electrons. 
Then, by increasing the energy $E$ and the incident angle $\phi$, the high transmission zones shrink while the low transmission gap extends for a large window of $V_{0}$ and $L$. In other words, as the barrier width increases, we observe a larger energy filtering effect, in addition to the overall decrease in transmission due to the increase in backscatter. This energy-filtering effect is useful, for example, in implementing low-noise devices \cite{Pan} or, more generally, in devices that require or exploit charge injection with a well-defined and limited amount of energy. When the barrier height $V_0$ is large enough, $V_{0}>\frac{3E}{2}$, the angular distribution of the transmission probability resembles that shown in Fig. \ref{fig6}a.  
In particular, in the classical forbidden zone, the transmission probability gets closer to the unity, which is the Klein tunneling. 

\begin{figure*}[tbph]
\begin{center}
\includegraphics[width=17.cm,height=11cm]{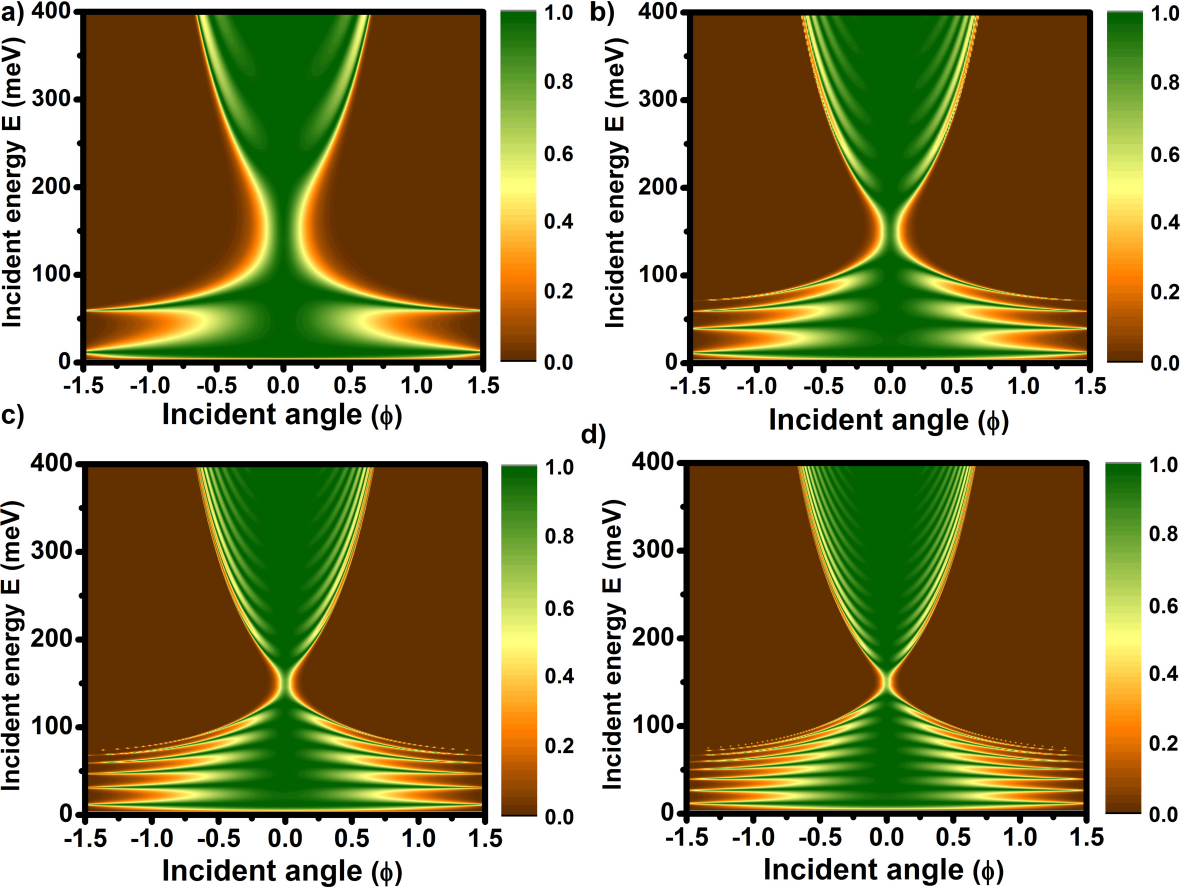}
\end{center}
\par
\vspace{-10pt}
\caption{(color online) Density plot of transmission probability $T$ as a function of the
incident angle $\protect\phi $ and and energy $E$  a
barrier height $V_{0}=150$ meV and different values of barrier width L:  a) $30$ nm, b) $60$ nm,  c)
$90$ nm, and d) $120$ nm.}
\label{fig7}
\end{figure*}

{Fig. \ref{fig7} shows a density plot of transmission probability $T$ as a function of incident angle $\phi $ and  energy $E$.}
%
%For a nonzero electrostatic potential $V_0$ applied in region $2$, the transmission probability mapping from $0$ (brown) to $1$ (olive green) as a function of incident energy $E$ and incident angle $\phi $ is shown in Fig. \ref{fig7}. 
Since there are no external influences, such as an applied magnetic field, the transmission is symmetric with respect to $\phi $, as can be clearly seen.
We observe a perfect transmission concentrated around the normal incidence angle $\phi = 0$, implying that the barrier is fully transparent to Dirac fermions (Klein tunneling \cite{1929}), regardless of  energy $E$. 
%Following Klein's discovery in 1932 of the tunneling of relativistic particles, this behavior is known as the Klein paradox \cite{1929}. 
As the energy rises from $E=0$, more resonant tunneling peaks may or may not arise depending on the incident angle since the incident electrons can enter the valence band. This is a manifestation of the Klein tunneling  \cite{Katsnelson}. The closer the energy value approaches the barrier height, the harder it will be for the electrons to cross the barrier region. When incident energy $E$ equals  barrier height $V_0$, the incident electrons must pass through the touch strip in the barrier region with the lowest density of state, and the majority of the incident electrons are reflected.  
However, when $E$  surpasses $V_0$, the transmission is significantly reduced, and the electrons are almost entirely reflected as the incident angle exceeds a critical value. Furthermore, we find a clear difference in the density plot of the transmission when the barrier width $L$ is considered. In fact, increasing $L$ significantly reduces transmission and increases the degree of fragmentation of the new minibands. 
As a result, new regions of low transmission probability and gaps appear, allowing the emergence of more tunneling minibands, which is due to the well-known Fabry-P\'{e}rot interference phenomenon \cite{fabry}. Moreover, it is noteworthy that with an increase in $L$, a greater number of resonances, typically occurring within the {potential barrier}, become compressed into narrower transmission zones. These results indicate a disruption in the symmetry of resonances, manifesting as a combination of resonance lines and ovals rather than the anticipated circular resonances. We mention that the asymmetry observed in Fig. \ref{fig7} is a consequence of the positive sign of the barrier width \(V_0\) because employing the mapping \(V_{0} \rightarrow -V_{0}\), Fig. \ref{fig7} will be reflected around this axis.

%%%%%%%%%%%%%%%%%%%%%%%%%%%%%%%%

\section{Conductance}\label{CON}

%%%%%%%%%%%%%%%%%%%%%%%%%%%%%%%%%%%%%%%%%%%%%%%%%%%%%%%%%%%
To capture the distinctive features of the transmission probability in a measurable quantity, we will depict the overall behavior of the effective conductance. At zero temperature, we have \cite{16}
\begin{equation}
G_{\text{eff}}=\frac{2\sqrt{E_{F}^{2}-\Delta _{so}^{2}}}{\hbar v _{F}\pi }%
\int_{\frac{-\pi }{2}}^{\frac{\pi }{2}}T(E_{F},V_{0},L,\phi )\cos \phi d\phi.
\label{22}
\end{equation}
Now if we express the widths in nm and the energies in meV, we get $\hbar v _{F}=658.2$ meV nm and the effective conductance in $e^2/h$ units. The conductance is determined by the potential applied by the top gate (via transmission), the potential applied by the back gate (via the Fermi energy $E_F$), and the width of the silicene sheet. Furthermore, \eqref{22} shows that the maximum value of effective conductance is at $T(E_{F},V_{0},L,\phi)=1$, and thus 
\begin{equation}
G_{\text{eff,max}}=\frac{4}{\hbar \upsilon _{F}\pi }\sqrt{E_{F}^{2}-\Delta _{so}^{2}%
}  \label{23}.
\end{equation}%
The effective conductance $G_{\text{eff}}$ for a the {potential barrier}   in silicene is obtained by substituting \eqref{19} into \eqref{22}. This process yields
\begin{widetext}
\begin{equation}
G_{\text{eff}}=\frac{2\sqrt{E_{F}^{2}-\Delta _{so}^{2}}}{\hbar \upsilon _{F}\pi }%
\int_{\frac{-\pi }{2}}^{\frac{\pi }{2}}\frac{\cos \phi d\phi }{1+\frac{%
V_{0}^{2}\left( \tan ^{2}\phi +\frac{(\Delta _{so}\sec \phi )^{2}}{\left(
E^{2}-\Delta _{so}^{2}\right) }\right) \sin ^{2}\left( \frac{L}{\hbar
\upsilon _{F}}\sqrt{V_{0}\left( V_{0}-2E\right) +\left( E^{2}-\Delta
_{so}^{2}\right) \cos ^{2}\phi }\right) }{\left( V_{0}\left( V_{0}-2E\right)
+\left( E^{2}-\Delta _{so}^{2}\right) \cos ^{2}\phi \right) }} . \label{24}
\end{equation}
\end{widetext}

\begin{figure*}[tbph]
\begin{center}
\includegraphics[scale=0.7]{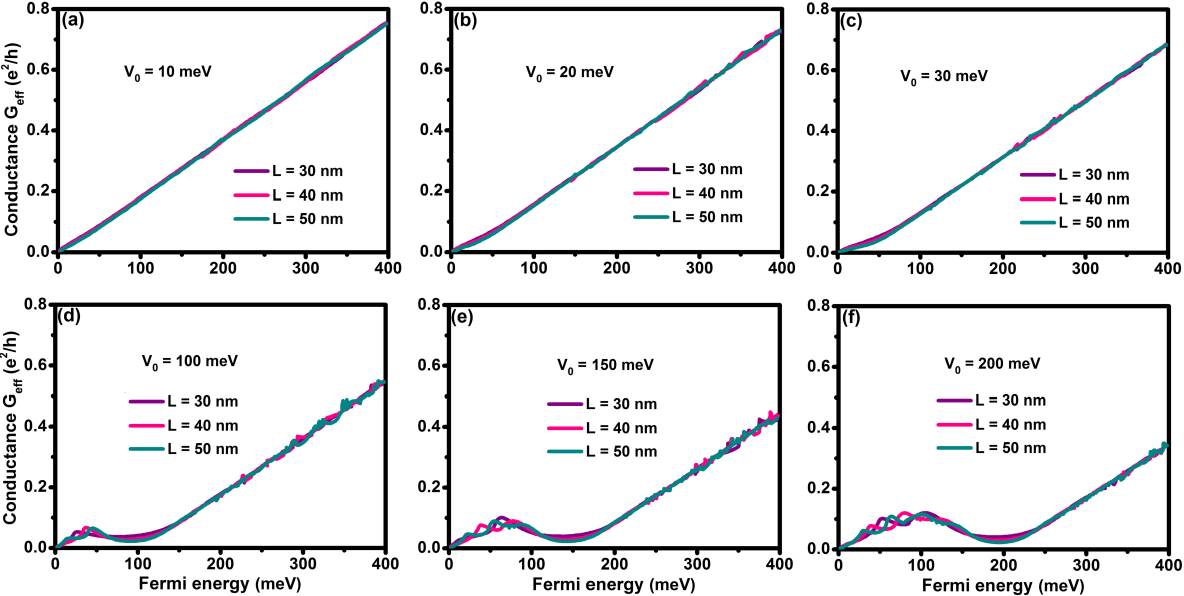}
\end{center}
\par
\vspace{-10pt}
\caption{(color online) Effective conductance $G_{\text{eff}}$ versus Fermi energy $E_F$ for different values of the
barrier height $V_0$ and width $L$.}
\label{fig8}
\end{figure*}

Our study of transport properties is supplemented by calculating the effective conductance $G_{\text{eff}}$ as a function of Fermi energy for various $ V_0 $ and $ L $ to reflect the striking differences between configuration-dependent transmissions in some quantifiable variables. The effective conductance $G_{\text{eff}}$ is shown in Fig. \ref{fig8} as a function of the Fermi energy $E_F$ for various $V_0$ and $L$ values.  $G_{\text{eff}}$ begins at zero for $E_F=0$ because the density of states outside the barrier vanishes at this energy. In the case where $V_0\leq 30$ meV, $G_{\text{eff}}$ increases with some oscillatory behavior that appears with increasing $V_0$.
However, when $V_{0}\geq 100$ meV, $G_{\text{eff}}$ displays a nonmonotonic dependence on the energy that differs considerably from that of $V_{0}\leq 30$ meV, leading to more complex behavior. Because of this non-monotonic variation, $G_{\text{eff}}$ increases initially with $E_{F}$ to reach a local maximum value when $E_{F}=\frac{V_{0}}{2}$. 
Then, $G_{\text{eff}}$ starts to decrease with $E_{F}$ as a result of the evolution of the density of states outside and inside the barrier. As a result, when the density of states inside the barrier vanishes at $E_F=V_0$, $G_{\text{eff}}$ reaches a nonzero minimum value. There are superimposed oscillations on this resonant conductance profile, with some local maxima and minima when $E_F<V_0$. Furthermore, tunneling effect transmission causes an increase in $G_{\text{eff}}$ for energies $E_F $ greater than $V_0 $. 
When $E_{F}=\frac{V_{0}}{2}$, the sign of the quantity $V_{0}(V_{0}-2E_{F})+(E_{F}^{2}-\Delta _{so}^{2})\cos ^{2}\phi$ changes from positive to negative. Therefore, in \eqref{24}, the sine of this quantity will become a hyperbolic sine, and $G_{\text{eff}}$ will become smaller. Thus, the resistance will increase. This situation will persist until a local minimum is reached at $E_{F}\simeq V_{0}$. 
When $E_{F}$ is greater than $V_{0}$, since the transmission coefficient tends to unity, $G_{\text{eff}}$ grows proportionally to the Fermi energy $E_F$, similarly to \eqref{23}. On the other hand, by varying the barrier width $L$, the overall behavior of $G_{\text{eff}}$ remains almost intact, which is consistent with what has been reported in the literature \cite{Deng,Jalil,Marconcini}.

\begin{figure*}[tbph]
\begin{center}
\includegraphics[scale=0.7]{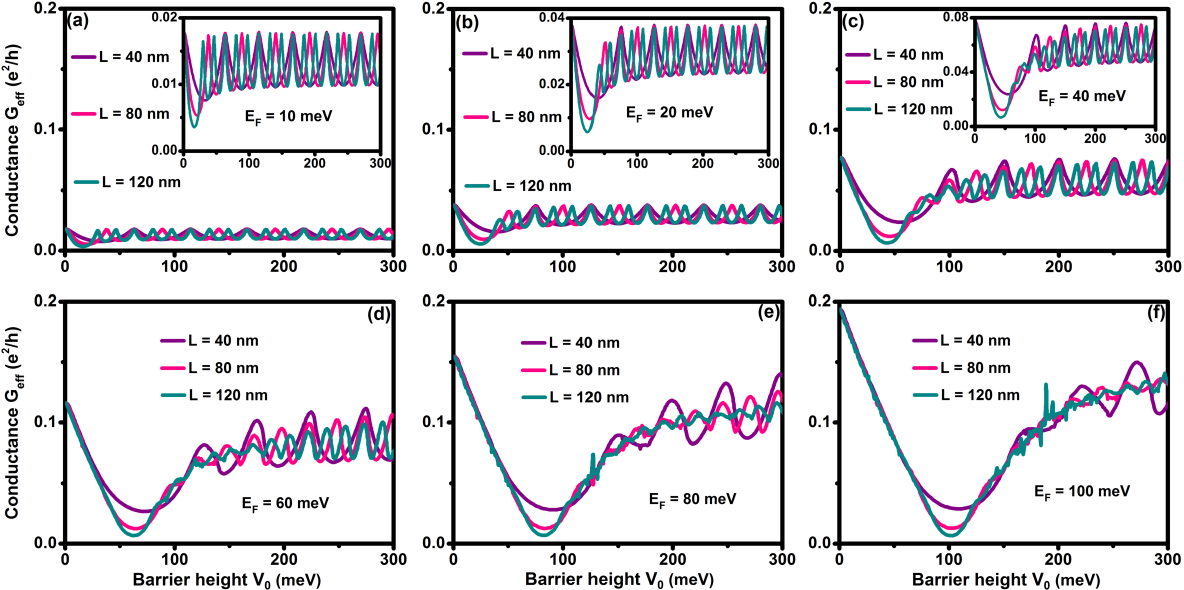}
\end{center}
\par
\vspace{-10pt}
\caption{(color online) Effective conductance $G_{\text{eff}}$ versus  the barrier height $V_0$ for different values of the Fermi energy $E_F$ and barrier width $L$.}
\label{fig9}
\end{figure*}

We mention that the transmission characteristics mentioned above can be directly mirrored from the quantifiable conductance spectrum shown in Fig. \ref{fig9}. It is clear that all conductance profiles are almost similar to each other and show non-monotonic variations as a function of barrier height $V_{0}$. We see that the Fermi energy $E_ F $ determines the onset of all spectra, with no significant differences as a function of barrier width $L$ when $V_ 0<E_ F $.
For $V_0=E_F$, the effective conductance $G_{\text{eff}}$ decreases monotonically as the barrier height $V_0$ increases to a smaller but still finite value. At this point, $G_{\text{eff}}$ reaches a local minimum that is dependent on the barrier width $L$ and approaches zero as $L$ increases. Because of the tunneling effect for $ V_0>E_F $, increasing $ V_0 $ would significantly increase $G_{\text{eff}}$ and gradually change the resonant nature of the profile to pronounced oscillatory behavior. In fact, it is noteworthy that all the obvious conductance oscillations shown in the insets for each case are due to the enhanced transmission through the classically forbidden regions. It is worth noting that as the barrier width $L$ decreases, the amplitude and period of the oscillations increase, whereas the rate of conductance increases as the Fermi energy $E_F$ increases. It is interesting to note that the conductance profiles here  are quite consistent with the results obtained in Fig. \ref{fig6}. These findings are consistent with those reported by Wang \cite{Wang}. When $V_0>E_F$, the barrier width has a significant impact on the transport properties, as shown by the comparison with the other results, particularly those discussed in Fig. \ref{fig8}. In particular, as the barrier width increases, we observe that the conductance oscillations are considerably affected.

\begin{figure*}[tbph]
\begin{center}
\includegraphics[scale=0.7]{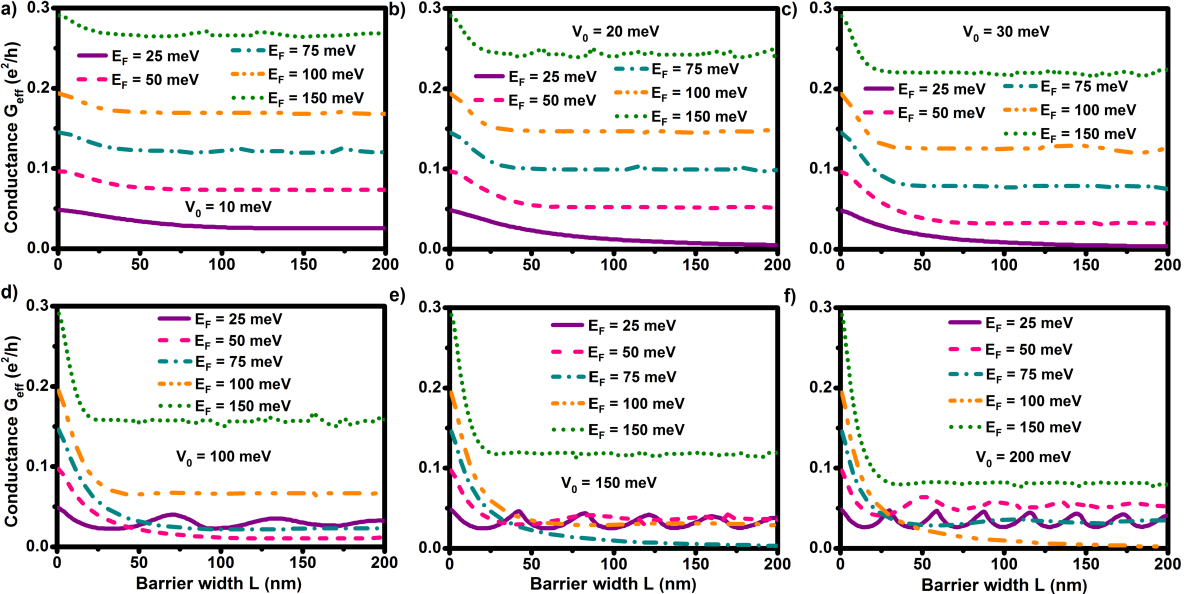}
\end{center}
\par
\vspace{-10pt}
\caption{(color online) Effective conductance $G_{\text{eff}}$ as a function of barrier width $L$ for different values of
barrier height $V_0$ and Fermi energy $E_F$.}
\label{fig10}
\end{figure*}

To shed more light on how the barrier width $L$ affects the effective conductance, we present Fig. \ref{fig10}. The conductance decreases monotonically with increasing barrier width $L$, as expected, due to the exponential decaying dependence of the transmission, as shown in Figs. \ref{fig10}(a,c). As a result, we note that the conductance becomes almost independent of the barrier width $L$. Such behavior is always maintained at a very low value and depends only on the barrier height $V_{0}$ and the Fermi energy $E_{F}$. Indeed, as $V_0$ increases, the rate of decrease of conductance increases, while it decreases as  $E_F$ increases. In another attempt, in Figs. \ref{fig10}(d,f), the qualitative nature of the conductance varies remarkably. Instead of monotonic stability, the resonant nature of the conductance profile for a low energy value changes significantly to an oscillatory one. Indeed, we observe that as the barrier height $V_{0}$ increases, the oscillating trend of the conductance increases and the number of oscillating peaks intensifies, as well as that the conductance of large values of energy exhibits more drastic drops to very low values. This study demonstrates that, in addition to Fermi energy $E_ F $ and barrier height $V_ 0 $, barrier width $L$ is a tunable parameter that can be used as a probe to learn more about the intrinsic properties of silicene. A similar behavior has been reported in multi-Weyl semimetals and phosphorene \cite{Deng,Biswas}.

%%%%%%%%%%%%%%%%%%%%%%%%%%

\section{Conclusion}

%%%%%%%%%%%%%%%%%%%%%%%%%%%%%%%%%%%%%%%%%%%%%%%%%%%%%%%%%%%
In this work, our focus is on predicting the remarkable quantum physical properties of silicene, characterized by its buckled honeycomb structure and the presence of an electric-tunable energy gap. Particularly, we placed special emphasis on elucidating its transport and tunneling properties. To model the behavior of carriers in silicene when scattered by a rectangular  {potential barrier}, we employed a Dirac-like Hamiltonian. Within this framework, essential physical quantities, including the transmission probability, reflection, and conductance, were derived.
% using the Landauer-B\"{u}ttiker formalism \cite{16}.

It is important to highlight that by imposing electrostatic manipulation of the rectangular {potential barrier}, the relevant transport properties can be efficiently engineered. This improvement in transmission quality and tunneling behavior represents an attractive feature for nanotransistor devices. The transmission probability was discovered to be extremely sensitive to barrier height $V_0$ and incident angle $\phi$, resulting in oscillatory behavior. The transmission profiles can also be controlled electrically.
% using Fermi energy $E_F$ control. 
In this context, we have investigated the tunneling features and the angular dependence of the transmission under the influence of barrier width $L$. Our calculations revealed several highly intriguing transmission characteristics, including perfect transmission and Fabry-P\'{e}rot resonances. It is intriguing that the perfect transmission regions are not restricted to normal incidence but may also be found at a range of oblique incidence angles. The Dirac electrons can therefore perfectly tunnel through a rectangular potential barrier. This phenomenon relates to the fact that even in the presence of high {potential barrier}s, there is no normal backscattering of electrons. Furthermore, when the barrier width increases, the transmission probability is considerably reduced, and the degree of fragmentation of the new minibands is increased. These tunnel minibands undergo the separation effect based on the Fabry-P\'{e}rot interference. Applications involving electro-optics can make use of these predictions.

We have also shown that the conductance characteristics of the system can be accurately controlled by simultaneously manipulating the Fermi energy and the {potential barrier}. This property can be investigated for the engineering of new nano-electronic circuits and semiconductor devices. Our findings showed that effective conductance behaves non-monotonically in relation to the Fermi energy $E_F$ and barrier height $V_0$. Interestingly, the local minimum conductance occurs when the Fermi energy and barrier height are equal. In addition, it is found that the conductance spectra display quantum oscillatory behavior with increasing Fermi energy and barrier height. On the other hand, we have shown that as the barrier width $L$ increases, the effective conductance decreases considerably until it approaches a constant value. For the thin barrier limit, the effective conductance is found to be independent of the barrier height. Our interpretation of the achieved results can serve as a foundation for a physical understanding of the transport mechanisms in transistors based on silicene.  This study can serve as a fantastic starting point for future theoretical and experimental studies in this area.

%\section*{Author Contributions}
%
%All authors contributed equally to this work. All authors have read and
%approved the published version of the manuscript.
%
%\section*{Declaration of competing interest}
%
%The authors declare that they have no known competing financial interests or
%personal relationships that might appear to influence the work presented in
%this paper.

%\section*{Acknowledgments}
%
%%%%%%%%%%%%%%%%%%%%%%%%%%%%%%%%%
%
%The generous support provided by the Saudi Center for Theoretical Physics
%(SCTP) is highly appreciated by all authors.

\end{document}